# Magnetoelectric coupling in multiferroic CoFe$_2$O$_4$/(Ba$_{0.95}$Ca$_{0.05}$)(Ti$_{0.89}$Sn$_{0.11}$)O$_3$ core–shell nanofibers elaborated by co-axial electrospinning method

Youness Hadouch[1,2,*], Daoud Mezzane[1,2], M'barek Amjoud[1], Valentin Laguta[3,4], Khalid Hoummada[5], Voicu Octavian Dolocan[5], Mustapha Jouiad[2], Mohammed Lahcini[1,6], Hana Uršič[7], Nikola Novak[7], Zdravko Kutnjak[7], Yaovi Gagou[2], Igor Lukyanchuk[2], Mimoun El Marssi[2].

1. *Laboratory of Innovative Materials, Energy and Sustainable Development (IMED), Cadi-Ayyad University, Faculty of Sciences and Technology, BP 549, Marrakech, Morocco.*
2. *Laboratory of Physics of Condensed Matter (LPMC), University of Picardie Jules Verne, Scientific Pole, 33 rue Saint-Leu, 80039 Amiens Cedex 1, France.*
3. *Institute of Physics AS CR, Cukrovarnicka 10, 162 53 Prague, Czech Republic.*
4. *Institute for Problems of Materials Science, National Ac. of Science, Krjijanovskogo 3, Kyiv 03142, Ukraine.*
5. *Aix-Marseille University - CNRS, IM2NP Faculté des Sciences de Saint-Jérôme case 142, 13397 Marseille, France.*
6. *Mohamed VI Polytechnic University, Lot 660, Hay Moulay Rachid. 43150 Ben Guerir Morocco.*
7. *Jozef Stefan Institute, Jamova Cesta 39, 1000 Ljubljana, Slovenia.*

**Abstract:**

Multiferroic CoFe$_2$O$_4$-Ba$_{0.95}$Ca$_{0.05}$Ti$_{0.89}$Sn$_{0.11}$O$_3$ core-shell nanofibers (CFO@BCTSn NFs) were synthesized by a sol-gel co-axial electrospinning technique. The scanning electron microscope and transmission electron microscope were used to check nanofibers' core-shell structure/configuration. X-ray diffraction and a high-resolution transmission electron microscope were used to confirm the spinel structure of CFO and the perovskite structure of BCTSn. The magnetic character of the resultant CFO@BCTSn NFs was determined by SQUID magnetometry. The piezoelectricity was verified using piezo-response force microscopy, which revealed an entirely covered ferroelectric shell outline, in accordance with SEM and TEM observations. The magnetoelectric (ME) coefficient was measured as a function of the applied external DC magnetic field. The maximum ME coefficient obtained for the CFO@BCTSn NFs was 346 mV cm$^{-1}$ Oe$^{-1}$. The high magnetoelectric coupling suggests that CFO@BCTSn NFs could be a promising candidate for magnetic field sensor and magnetoelectric device applications.

## Declarations

*Conflicts of interest/Competing interests*

*Not applicable*

*Availability of data and material*

*Not applicable*

*Code availability*

*Not applicable*

*Ethics approval*

*Not applicable*

*Consent to participate*

We confirm that all authors mentioned in the manuscript have participated in, read and approved the manuscript, and have given their consent for the submission and subsequent publication of the manuscript.

*Author Contribution*

*All authors certify that they have participated sufficiently in the work to take public responsibility for the content. Furthermore, each author certifies that this work will not be submitted to other journal or published in any other publication before.*

Y. Hadouch: Investigation, Methodology, Data Curation, Writing Original Draft, Validation.

D. Mezzane: Visualization, Methodology, Writing - Review & Editing, Validation, Supervision.

M. Amjoud: Visualization, Writing - Review & Editing, Validation, Supervision.

V. Laguta: Writing - Review & Editing, Visualization, validation,

K. Hoummada: Review & Editing, Visualization, validation,

V. Dolocan: Writing - Review & Editing, Visualization, validation,

M. Jouiad: Review & Editing, Visualization, validation,

M. Lahcini: Review & Editing, Visualization, validation,

H. Uršič: *Writing - Review & Editing, Visualization, validation,*

N. Novak: Review & Editing, Visualization, validation,

Z. Kutnjak: Writing - Review & Editing, Visualization, validation,

Y. Gagou: Formal analysis, Writing, validation, Supervision.

I. Lukyanchuk: Review & Editing, Visualization, validation,

M. El Marssi: Visualization, Validation, Writing - Review & Editing, Supervision.

*Consent for publication*

*We confirm that all the authors mentioned in the manuscript have agreed to publish this paper.*

# 1. Introduction

Magnetoelectric materials have attracted much attention in recent years [1], [2]. The coupling between the electric and the magnetic ordering, they can convert an electric field into a magnetic field and vice versa via the mechanical interactions between the ferroelectric and ferromagnetic phases [3]. This opens the door to a wide range of technological applications, such us multiple-state memory devices that can be written electrically and read magnetically, electrically controlled microwave phase shifter or voltage-controlled ferromagnetic resonance, magnetically controlled electro-optic or piezoelectric devices and magnetoelectric memory cells [4–8].

There are two types of magnetoelectric materials: (i) single-phase multiferroic in which magnetic and electric fields are intrinsically coupled [9], [10]. However, magnetoelectric couplings in this type of materials are typically very weak [11]. (ii) Multi-phase composites with spatially separated piezoelectric and magnetostrictive phases connected through an interface, where the magnetoelectric coupling is induced indirectly via a strain interaction between piezoelectric and magnetostrictive effects [12–23]. The significant factor in achieving strong ME effects in these composites is to transfer as much strain as possible from one phase of the composite to another [2], [24]. As a result, the area of the interface between ferroelectric and ferromagnetic grains is critical in this type of strain-mediated ME coupling [22], [25–28].

Many approaches and designs have been investigated to create hybrid multiferroic composites, called connectives, including 0–3, 1–3 and 2-2 type connectivity [23], [29]. The most studied type of connectivity is 0-3, also known as particulate composites [29–36]. In this type, the magnetic particles are enclosed by the piezoelectric phase. Specifically, these composites are made by combining the piezoelectric and magnetostrictive phases, resulting in inhomogeneous dispersion of the magnetic phase and agglomeration of these particles due to their attractive nature and surface energy. Agglomeration reduces the interface surface between the two ferroic phases, lowering the ME effect [23]. Based on this, the best way to intimate the contact between the two phases is to create one-dimensional (1D) interfaces at the nanoscale [38], [39]. The co-axial core-shell nanofibers, with a large interfacial surface and high surface-to-volume ratio, are the most commonly recommended for meeting these requirements [25–27], [39]. Several techniques, such as template-assistant approaches, hydrothermal methods, and electrospinning techniques, have been used to create 1D multiferroic nanocomposites [41], [42]. In particular, electrospinning is a feasible and adaptable method for creating multiferroic composite NFs with ultra-long and continuous structures. It has many advantages in practice, such as the ability to control the size of the nanostructures, the high aspect ratio, and the low cost-efficiency. Additionally, it is possible to easily control the properties of nanofibers using electrospinning parameters like the electric field, solution viscosity, humidity, and annealing temperature[43], [44]. On fundamental level the study of the multiferroic nanofibers presents a particular interest. Exotic polar topological states, like vortex and helix polarization tubes were predicted but not yet confirmed experimentally [45]–[49]. Understanding of coupling of such topological formations with magnetism is on track. In a theoretical study published 13 years ago, Zhang et al. showed that nanofiber composites have orders of magnitude stronger magnetoelectric responses than multiferroic composite thin films of similar compositions [1]. Despite all these benefits, very few studies have reported the fabrication of nanostructured composite fibers using a sol-gel assisted electrospinning technique. Furthermore, a small portion of these studies has focused on the magnetoelectric coupling of electrospun fibers [27], [43], [44]. These examples used cobalt ferrite $CoFe_2O_4$ (CFO) as a magnetostrictive phase with a large magnetoelastic coefficient (greater than 200 ppm) [52]–[55]. Due to its significant piezoelectric properties, $PbZr_{0.52}Ti_{0.48}O_3$ (PZT) is the ferroelectric material most frequently used in these composites [21], [22], [27], [56]. Unfortunately, Pb-based material is restricted globally because of its negative impact on the environment. Barium titanate $BaTiO_3$ (BT) and its derivatives are therefore the main candidates for a Pb-free multiferroic composite.

We have recently reported results obtained on $(1-x)Ba_{0.95}Ca_{0.05}Ti_{0.89}Sn_{0.11}O_3–xCoFe_2O_4$ particulate composites, and the maximum ME coefficient has been found around 0.1 mV cm$^{-1}$ Oe$^{-1}$ at 2.1 kOe for x=0.3 [57]. This value, while comparable to similar systems, remains low for such a multiferroic application. As reported in this work, the efficiency of ME coupling in BCTSn-CFO ceramic composites can be significantly improved by varying the sizes and shapes of the grains of the piezoelectric and magnetostrictive phases, as well as the synthesis of core-shell nanostructures.

This paper focuses on (i) the synthesis of magnetoelectric core-shell nanofibers with a $CoFe_2O_4$ core and a $Ba_{0.95}Ca_{0.05}Ti_{0.89}Sn_{0.11}O_3$ shell using the electrospinning method and (ii) the investigation of the structural, morphological, magnetic, ferroelectric and magnetoelectric properties of the CFO@BCTSn NFs.

## 2. Experimental section

### 2.1. Material synthesis

CFO@BCTSn nanofibers were prepared by sol-gel process based electrospinning method as follows. For CFO solution, 1 mmol of cobalt nitrate hexahydrate (Co(NO$_3$)$_2$.6H$_2$O) and 2 mmol iron nitrite nonahydrate (Fe(NO$_3$)$_3$.9H$_2$O) were dissolved into a mixed solution of 2 mL of ethanol and 2 mL of N,N-dimethyl formamide (DMF) and stirred continuously for 3 h. BCTSn precursor solution was synthesized by dissolving a stoichiometric portion of barium acetate Ba(CH$_3$COO)$_2$ and calcium acetate Ca(CH$_3$COO)$_2$ in acetic acid and tin chloride (SnCl$_2$.2H$_2$O) in 2-methoxyethanol separately. The polymer solution was prepared by dissolving poly(vinyl pyrrolidone) (PVP) with a molecular weight of 1 300 000 in ethanol under vigorous stirring for 4 h. The CFO and BCTSn solutions were added to the PVP solution and stirred continuously to form homogenous CFO and BCTSn polymer solutions, with the concentration of PVP controlled around 0.045 and 0.030 g mL$^{-1}$, respectively. Individual polymer solutions were transferred into glass syringes for electrospinning. The coaxial spinneret and electrospinning setup are depicted in **Fig. 1**. The spinning process was performed at DC voltage with a 12 cm distance between the needle tip and collector. The BCTSn and CFO polymer solutions were pumped at a rate of 0.5 and 0.3 mL h$^{-1}$, respectively. The as-spun NFs were dried at 80 °C under vacuum for 12 h before being annealed at 700 °C for 4 h in an air atmosphere.

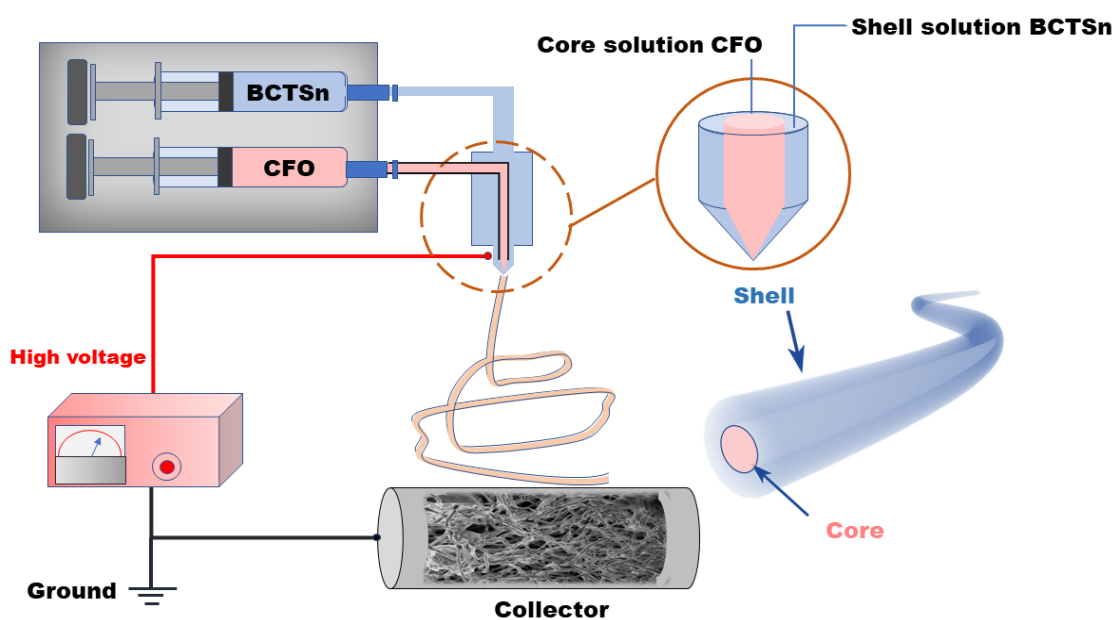

**Fig. 1**. Coaxial electrospinning of core-shell nanofibers: setup schematics.

### 2.2. Characterizations

The core-shell nanofibers microstructure was analyzed using scanning electron microscopy (SEM) in a HELIOS 600 nanolab from Thermofisher and a (JEOL – ARM200F Cold FEG) high resolution analytical transmission electron microscope (TEM) operating at 200KV. The XRD patterns of CFO@BCTSn NFs were obtained by X-ray diffraction in the Bragg-Brentano (θ-2θ) geometry using a Cu Kα source ($\lambda K\alpha = 1.54$ Å) at room temperature. The local piezoelectric responses of CFO@BCTSn NFs were investigated by an atomic force microscope (AFM, Jupiter XR Asylum Research, Oxford Instruments, CA, USA) equipped with a piezo-response force module (PFM). Pt-coated silicon tips with a radius of curvature ~10 nm (OMCL-AC240TM-R3, Olympus, Japan) were used for the PFM analysis. To prevent the CFO@BCTSn NFs from sticking to the PFM tip, the fibers were fixed to the Si substrate by heating at 600 °C for 30 min. We chose the nonconductive Si substrate to avoid electrical breakdown through the air and short-circuiting between the tip and the conductive substrate when scanning over the fiber edge and touching the substrate with the PFM tip. Measurements were performed in virtual ground mode as described in our previous work [58], [59]. The images were scanned in dual AC resonance-tracking lateral (DART Lateral) mode. An electrical voltage of 4 V and a frequency of ~690 kHz was applied. Magnetic properties were measured by demonstrating a magnetic hysteresis (M-H) loop at room temperature using a SQUID magnetometer Quantum Design MPMSXL under a magnetic field range of 0–50 kOe. For magnetoelectric measurements, a conventional Bruker EPR spectrometer EMX plus was used. For this measurement, a rectangular sample with a dimension of 3×5×0.3 mm$^3$ was prepared by bonding the nanofibers with epoxy resin. Before the measurement, the sample was coated with silver electrodes and poled at room temperature by applying a dc electric field of 10 kV/cm for 10 min (More details are described in Ref. [60]).

## 3. Results and discussion

**3.1 CFO@BCTSn NFs: morphology and structure:**

The morphologies of the annealed CFO@BCTSn NFs were observed by SEM and TEM. **Fig. 2a** depicts an SEM image of the CFO@BCTSn NFs, demonstrating that individual NFs have a continuous structure with diameters ranging from 150 to 300 nm. The surface of a single CFO@BCTSn NF is shown in **Fig. 2b**, indicating that the fiber is composed of small grains in the range of tens of nanometers, that may be of the BCTSn phase because they form the fiber's shell. A detailed observations of an individual fiber are depicted in **Figs. 2c and d**. The core and shell are well identified, with a 110 nm diameter CFO core and a 370 nm thick BCTSn shell.

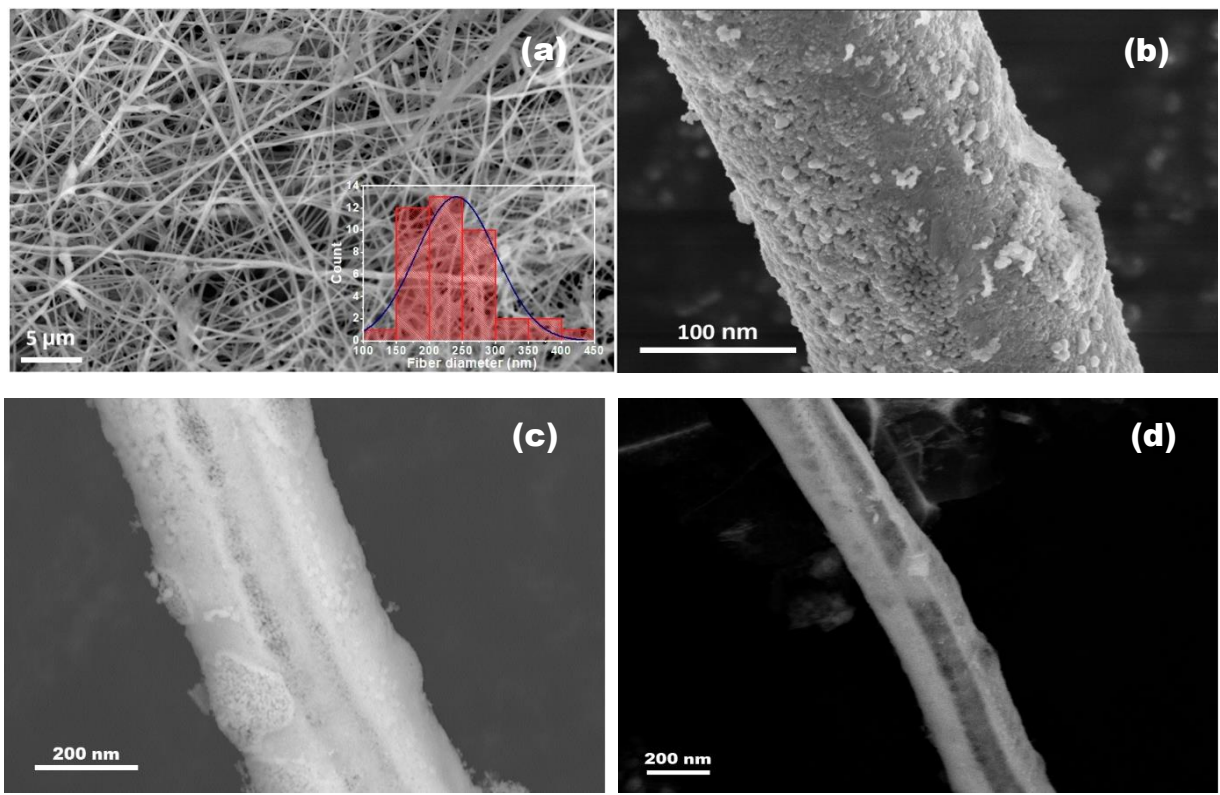

**Fig. 2.** SEM images of annealed (a) CFO@BCTSn NFs, the inset to the image shows a fiber diameter distribution, (b), (c) and (d) single CFO@BCTSn NF.

The microstructures of the CFO@BCTSn NFs were further investigated by TEM. **Fig. 3a** depicts a TEM image of an individual nanofiber, made up of nanocrystallites of CFO and BCTSn that are connected individually. **Fig. 3b** depicts the interference of two crystallographic orientations originating from BCTSn and CFO crystals. The SAED pattern in the inset of **Fig. 3a** obtained from the red colored-circle zone shows multiple diffraction rings, suggesting a polycrystalline structure of CFO@BCTSn NFs. The crystallographic planes (400), (440), (422) and (620) are assigned to CFO [61], while the planes (220) and (223) are assigned to BCTSn [62], confirming the coexistence of both magnetic and ferroelectric phases.

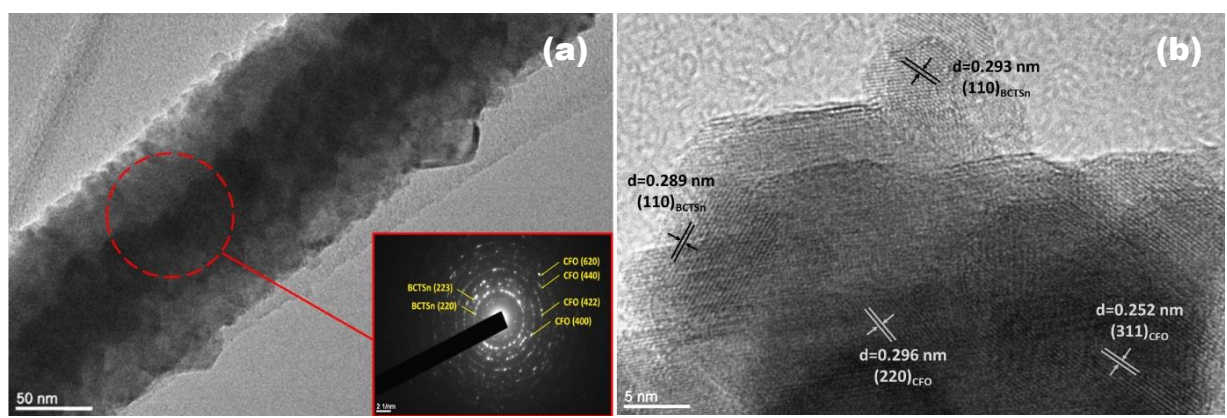

**Fig. 3**. (a) TEM image of an individual CFO@BCTSn NF, inset in (a): diffraction pattern acquired from the red-colored circle, (b) HRTEM image of CFO@BCTSn NF.

The XRD data of CFO@BCTSn NFs are shown in **Fig. 4a**, which reveals two types of diffraction peaks that are consistent with the Spinel structure's standards (JCPDS No. 01-1121) and the perovskite structure (JCPDS No. 31-0174). Furthermore, the average crystallite size of CFO and BCTSn is estimated to be 4.4 and 5.1 nm, respectively, based on a detailed analysis of the peaks broadening of (311) and (100) reflections using the Debye-Scherrer formula[62].

The lattice strain in the CFO@BCTSn NFs is calculated using the Williamson-Hall (W-H) method [21]. The lattice strain of the composite is calculated by considering the peaks in the XRD patterns that correspond to BCTSn, as shown in **Fig. 4a**.

$$\beta \cos\theta = 4\varepsilon \sin\theta + \frac{k\lambda}{D} \quad (1)$$

where k (=0.9) is the shape factor, λ is the wavelength of X-ray, θ is the peak position in degree, D is the average crystallite size, ε is the lattice strain, and β is the full width at half maxima (FWHM). According to equation (1), the plot of β cos θ versus 4sin θ should be a linear function, with the slope indicating the strain in the sample. The lattice strain in CFO@BCTSn NFs is $3.64 \times 10^{-3}$. This positive value indicates that the strain in the CFO@BCTSn NFs is tensile [21]. It is well known that lattice strain improves polarization stability and piezoelectric properties in the ferroelectric phase of magnetoelectric composite materials [63], [64].

These obtained results confirm the successful preparation of the 1D CFO@BCTSn NFs multiferroic composite material. The multiferroic properties of the CFO@BCTSn NFs are then investigated by measuring their ferromagnetic, ferroelectric and ME coupling.

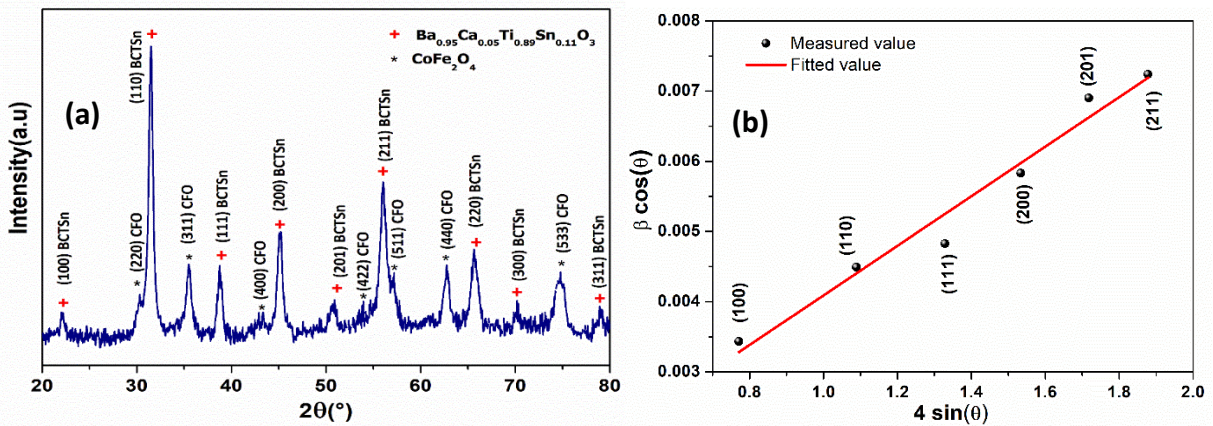

**Fig. 4.** (a) Room temperature XRD pattern and (b) Williamson-Hall (W–H) plot of CFO@BCTSn NFs.

## 3.2 CFO@BCTSn NFs magnetic properties:

The magnetization vs. magnetic field for CFO@BCTSn NFs at 300 K is shown in **Fig 5**. The M-H loop shows a classical ferromagnetic behavior with a very small paramagnetic component due to the possible interdiffusion of a small number of magnetic atoms in the shell. The inset displays a zoom of the area depicted by the rectangle. We determine a coercive field of 221 Oe, a remanent magnetization $M_r$ of 1.43 emu g$^{-1}$ (only 12 % of the maximum value) and a maximum magnetization $M_s$ of 11.63 emu g$^{-1}$ at room temperature. It should be stressed that the magnetization is not completely saturated under 25 kOe.

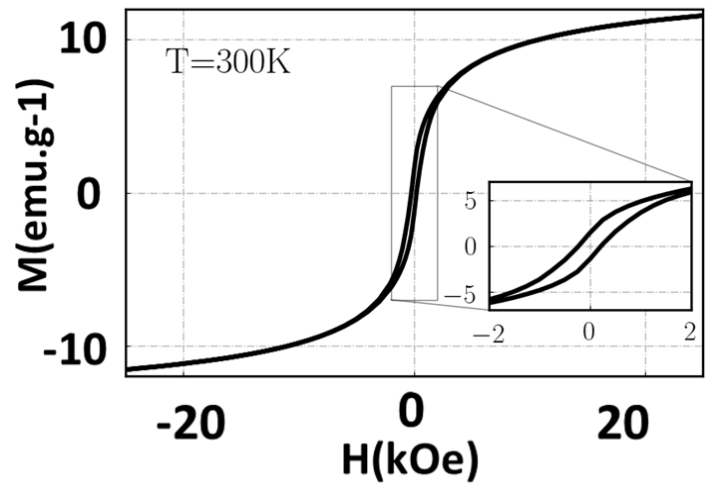

**Fig. 5.** Magnetic hysteresis loops for the CFO@BCTSn NFs at room temperature.

## 3.3 Ferroelectric properties of CFO@BCTSn NFs:

The lateral PFM images of the nanofibers on the Si substrate are shown in **Fig. 6**. Comparing the topography panels (a) and (b) with the PFM panels (c) and (d), it is clear where the nanofibers are positioned on the Si substrate. The nanofibers show enhanced piezoelectric response, which can be seen as bright areas in panel (c), except in the center of the nanofibers, where there is no piezoelectric signal (the dark spots). Some examples of areas without piezoelectric signals are marked with blue arrows. These areas most likely correspond to the magnetic phase inside the nanofibers.

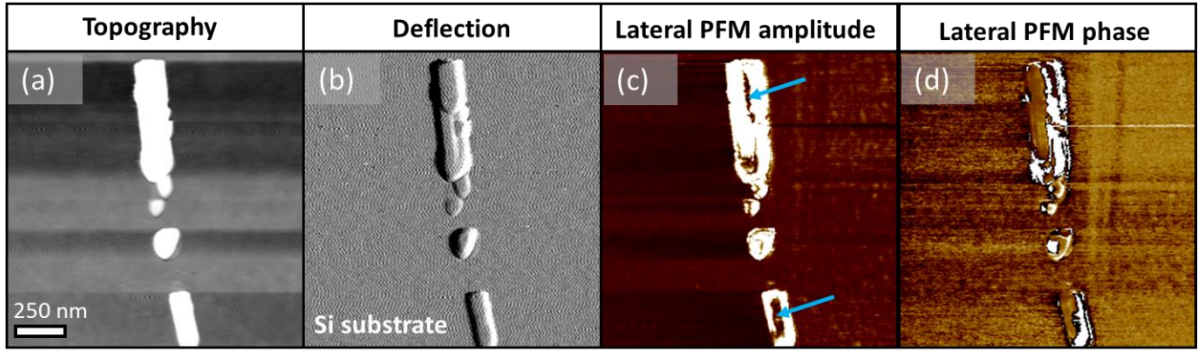

**Fig. 6.** AFM/PFM images of CFO@BCTSn NFs: AFM topography (a) height and (b) deflection and lateral PFM (c) amplitude and (d) phase images.

### 3.4 Magnetoelectric properties of CFO@BCTSn NFs:

The ME coupling coefficient between the BCTSn and CFO phases was calculated using the following formula described in our previous study [57].

$$\alpha_{ME} = \left(\frac{\partial E}{\partial H}\right) = \frac{V}{d\, H_{ac}} \quad (2),$$

where $V$ is the induced ME voltage across the sample, $d$ represents the thickness of the sample, and $H_{ac}$ is the applied AC magnetic field.

The room temperature ME coefficient ($\alpha_{ME}$) of the CFO@BCTSn NFs has been measured by applying the AC magnetic field of 1.78 Oe at 564 Hz frequency. The ME coefficient shows a non-linear increase with increasing dc magnetic field, reaching a maximum value of 346 mV cm$^{-1}$ Oe$^{-1}$ at the DC magnetic field of 10 kOe, as shown in **Fig. 7**. Because of the mechanical interactions between the ferrite and ferroelectric/piezoelectric perovskite phases throughout the fiber, the magnetic field induces strain in the ferrite phase, which results in stress in the ferroelectric phase. This stress produces a voltage in the shell via the piezoelectric effect [33].

As discussed in our previous paper [57], the area of the interface between ferroelectric and ferromagnetic phases and phase connectivity are crucial in the strain-mediated ME composites. The improved magnetoelectric coefficient obtained in the CFO@BCTSn NFs directly relates to electrospinning process. Indeed, the CFO@BCTSn NFs produced by electrospinning comprise nanoscale BCTSn and CFO grains as shown in SEM images (**Fig. 2**). Such nanoscale grains can make a large interfacial area, favoring mechanical interactions between the piezoelectric and magnetostrictive phases. In addition, our composite fibers also have good phase-connectivity, as observed in the TEM images (**Fig. 3**). This can ensure complete strain transfer between the CFO and BCTSn phases, enhancing the magnetoelectric properties of these composites. As a result, the observed $\alpha_{ME}$ is higher than the $\alpha_{ME}$=13.3 mV cm$^{-1}$ Oe$^{-1}$ reported for BaTiO$_3$/CoFe$_2$O$_4$ composite fibers [65], $\alpha_{ME}$=150.58 mV cm$^{-1}$ Oe$^{-1}$ reported for CFO@BT@PDA/(P(VDF-TrFE) and $\alpha_{ME}$=51 mV cm$^{-1}$ Oe$^{-1}$ reported for CoFe$_2$O$_4$ nanorods embedded in BaTiO$_3$ matrix [50]. However, the obtained $\alpha_{ME}$ value is lower than 2.95 10$^4$ mV cm$^{-1}$ Oe$^{-1}$ reported for CoFe$_2$O$_4$–Pb(Zr$_{0.52}$Ti$_{0.48}$)O$_3$ core-shell nanofibers [27], which may be attributed to the d$_{33}$ piezoelectric coefficient of Pb(Zr$_{0.52}$Ti$_{0.48}$)O$_3$ that is three or four times higher than that of the co-doped BaTiO$_3$ [51].

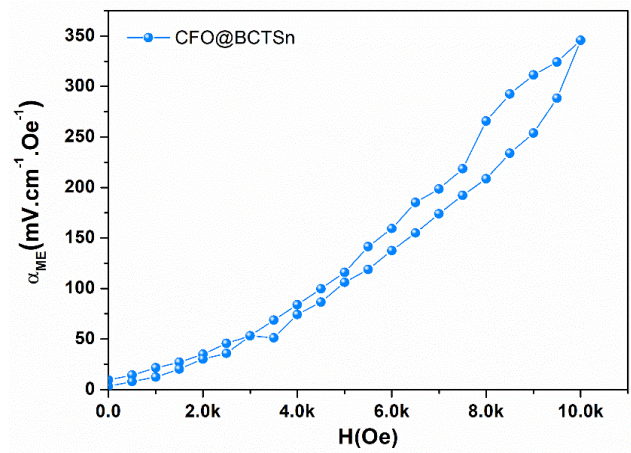

**Fig. 7**. Dependence of the magnetoelectric coefficient of CFO@BCTSn NFs on the DC magnetic field at room temperature.

## 4. Conclusion:

In this study, core-shell CoFe$_2$O$_4$-Ba$_{0.95}$Ca$_{0.05}$Ti$_{0.89}$Sn$_{0.11}$O$_3$ nanofibers were elaborated by the sol-gel electrospinning method. The structural, morphological, and multiferroic properties of these nanofibers were investigated. The presence of perovskite and spinel structures in the CFO@BCTSn NFs was confirmed by XRD and SAED analysis. The morphology of the core-shell nanofibers was observed by SEM and TEM images that endorse the core-shell connectivity. Multiferroic properties of CFO@BCTSn NFs were verified. PFM mappings and M-H hysteresis loops were used to prove that the synthesized fibers had both piezoelectric and ferromagnetic properties. The magnetoelectric coupling coefficient of 346 mV cm$^{-1}$ Oe$^{-1}$, determined experimentally, is significantly higher when compared to the other connectivity types. The high ME coupling was explained by the core-shell fibers' connectivity, which has large interfacial areas and excellent interconnectivity between the two ferroic phases.

**Acknowledgements:**

This research is financially supported by the European Union Horizon 2020 Research and Innovation actions MSCA-RISE-ENGIMA (No. 778072), MSCA-RISE-MELON (No. 872631). H.U acknowledge the Slovenian Research Agency (N2-0212, P2-0105) and thanks J. Cilenšek and V. Fišinger for help in the laboratory. ZK acknowledge the Slovenian Research Agency support (P1-0125).